\documentclass[aps,twocolumn,showpacs,amsmath,amssymb,pra,superscriptaddress,floatfix,longbibliography]{revtex4-1}

\usepackage{bm}% bold math
\usepackage{physics}
\usepackage{braket}
\usepackage{amsmath}
\usepackage{amssymb}
\usepackage{mathtools}
\usepackage{graphicx}% Include figure files
\usepackage{dcolumn}% Align table columns on decimal point
\usepackage{bm}% bold math
\usepackage{multirow}
\usepackage{leftidx}
\usepackage{color}
\usepackage{float}
\usepackage{amsfonts}
\usepackage[capitalise]{cleveref}
\usepackage[german,english]{babel}
\graphicspath{ {./Images/Free}, {./Images/bound} }
\begin{document}

%\preprint{}

\title{Simulation of nonlinear Compton scattering from bound electrons}


\author{Akilesh Venkatesh}
\affiliation{Department of Physics and Astronomy, Purdue University, West Lafayette, Indiana 47907, USA}

\author{Francis Robicheaux}
%\thanks{}
\affiliation{Department of Physics and Astronomy, Purdue University, West Lafayette, Indiana 47907, USA}

\date{\today}

\begin{abstract}
Recent investigations by Fuchs et al. [Nat. Phys. 11, 964 (2015)] revealed an anomalous frequency shift in non-linear Compton scattering of high-intensity X-rays by electrons in solid beryllium. This frequency shift was at least 800 eV to the red of the values predicted by analytical free-electron models for the same process. In this paper, we describe a method for simulating non-linear Compton scattering. The method is applied to the case of  bound electrons in a local, spherical potential to explore the role of binding energy in the frequency shift of scattered X-rays for different scattered angles. The results of the calculation do not exhibit an additional redshift for the scattered X-rays beyond the non-linear Compton shift predicted by the free-electron model. However, it does reveal a small blue-shift relative to the free electron prediction for non-linear Compton scattering. The effect of electron-electron correlation effects is calculated and determined to be unlikely to be the source of the redshift. The case of linear Compton scattering from a photoionized electron followed by electron recapture is examined as a possible source of the redshift and ruled out.
\end{abstract}

%\keywords{}

\maketitle
\section{Introduction}
Since, the discovery of Compton scattering 90 years ago, various measurements have been carried out to confirm the results to a higher accuracy and to probe the finer details of the Compton spectrum~\cite{postcompton1,postcompton2}. This has given rise to the study of Compton profiles which provide extensive information about the momentum distribution of the electrons involved in 
the scattering~\cite{Comptonprofile,Comptonprofile2}. Compton profiles have also proven useful as an experimental check on the accuracy of the ground state wave function of electrons in momentum space obtained through theoretical methods. A number of applications in areas from material science to astrophysics have been born out of these studies~\cite{Comptonapplication1,Comptonapplication2}.

In this paper, non-linear Compton scattering refers to the process where two incoming photons interact with an electron leading to one outgoing photon. Non-linear Compton scattering was first described by Brown and Kibble~\cite{KB} in 1964 where they developed an analytical QED framework to model the non-linear scattering~\cite{Vachaspati} of photons by a free electron. In their work, they showed that when non-linear Compton scattering occurs, for the non-relativistic case, the frequency of the scattered photon can be obtained by the usual Compton expression, provided, you replace the incoming frequency by twice that value. Including relativistic effects in the calculation gives rise to ponderomotive forces on the electron. At extremely high intensities (electric field $>$10000 a.u. for X-rays), the ponderomotive effects lead to the electron behaving as if it had a smaller mass and thus producing a bigger redshift for the scattered photons. It was more than two decades before experiments could study non-linear X-ray-matter interactions, but the arrival of X-ray free-electron lasers~\cite{XFEL1,XFEL2} has made considerable progress~\cite{NLXFEL1,NLXFEL2} possible.

More recently, Fuchs et al.~\cite{Fuchs} carried out an experiment to investigate non-linear X-ray matter interactions with the Linac Coherent Light Source at the SLAC National Accelerator Laboratory. They used a high-intensity X-ray free-electron laser to study non-linear scattering from solid beryllium. While non-linear Compton scattering had been earlier observed~\cite{firstNLC1}, Fuchs et al.~\cite{Fuchs} found a non-linear Compton signal that was substantially redshifted from the value predicted by Brown and Kibble ~\cite{KB}. To explain this additional redshift($\sim$800 eV), they proposed that the bound nature of the beryllium electrons could be responsible. This argument was analyzed by Krebs et al.~\cite{Santra}. They solved the TDSE to simulate the non-linear X-ray scattering, with the bound electrons being modelled by a potential based on the Hartree-Fock-Slater model. Their calculations did not reveal any anomalies with respect to the free-electron results.

In this paper, we re-examine the additional frequency shift in Fuchs et al.~\cite{Fuchs}. We use a numerical approach different from that of Ref.~\cite{Santra} and study the effect of binding energy, electron-electron correlation, and photo-ionization on the non-linear Compton spectrum. We were able to obtain convergent results for both the differential cross section and the average scattered photon momentum for both linear Compton and non-linear Compton scattering. While we mainly agree with the results of Krebs et al.~\cite{Santra}, our calculations reveal a small blue-shift in the frequency of the scattered photon with respect to the free-electron results. Following this, we explore two possible alternate causes for the redshift. First, we consider the role of electron-electron correlation effects on the scattering profile. Second, we examine the possibility of a semi-Compton process to give rise to the anomalous redshift.

For free-electrons, we performed calculations where the electron part of the wave function was restricted to 2D but for most of the bound electron calculations, the electron was fully 3D. For a given number of dimensions, the calculations for a bound electron involves less time and space computationally than their free-electron counterparts. It should be noted that a 2D model is quite adequate to describe both linear and non-linear Compton scattering but the exact factors required to calculate the differential cross length is not well defined. The 3D simulations lead to results with no adjustable parameters.

The model's validity is demonstrated by reproducing the differential cross length of X-ray scattering from a free-electron from a QED-2+1 scheme, which is a 2D analogue of the Klein-Nishina formula~\cite{QED2plus1,KN}. Another way the validity of the model is tested is by comparing it with the non-linear Compton differential cross section of Brown and Kibble ~\cite{KB} for small binding energy. Finally the model is applied to the X-ray scattering scenario in Fuchs et al.~\cite{Fuchs} to study both the Compton and non-linear Compton scattering from a bound electron. In our calculations, we consider a range of binding energy for the bound electrons from 0.4 a.u. to 6 a.u. This range of binding energy is relevant for Be because, the atomic Be has an ionization potential of 0.34 a.u. and that of Be$^{2+}$ is 5.6 a.u.

Unless otherwise stated, atomic units will be used throughout this paper. 

\section{Methods and Modelling}\label{Methods}

The first step in our approach is to model the initial state of the electron. For the free-electron case, we use a Gaussian wave-packet as the initial state.
Recently, Pan and Gover~\cite{initialwfn} while analyzing spontaneous and stimulated emissions found that the size of the initial wave packet has non-trivial effects on the spectrum of the outgoing photons. These effects appear when the outgoing photons are in a coherent state and not a Fock state. However for the scattering problem under consideration, the size of the Gaussian wave packet is not significant.

For the bound electron case, we treat the electron as an atomic single electron and model the rest of the atom with an effective time-independent, local potential.  We solve the time-independent Schr\"odinger equation to obtain the ground-state spatial wave function. For this, we use the relaxation method; propagating the Schr\"odinger equation in imaginary time until only the ground state remains. The ground-state wave function, thus obtained, was the initial state of the bound electrons in our calculations. 

With the appropriate initial wave function, we can compute the time-dependent wave function for the electron in a classical field by numerically solving the time-dependent Schr\"odinger equation(TDSE). To model the scattered photon, we employ lowest order perturbation theory and solve for the case of a single outgoing photon. We obtain the scattering probability for different angles which is used to calculate the differential cross section as a function of angle.

The non-relativistic treatment of the electron implied by the TDSE should be accurate enough for the conditions below. Consider the case of non-linear Compton scattering of a photon of $w=340$ a.u. from a free electron. Even for the case of back scattering, the electron would at most gain approximately 1.2 keV of energy from the photon [Eq. (\ref{NLC_freq})]. From the experiment by Fuchs et al.~\cite{Fuchs}, we expect an additional kinetic energy gain of approximately 1 keV. Together, that would still give a Lorentz factor($\gamma$) of only ~1.004 which is well within the non-relativistic regime. As a check on the approximation, we consider the lowest order relativistic correction to the Schr\"odinger equation in Sec. \ref{Rel-correction} and demonstrate that it hardly changes the overall results.

\subsection{Deriving the non-homogeneous Schr\"odinger equation} \label{derivingnon-homogenoeous}
We model the vector potential by treating the incoming EM wave classically and quantizing the scattered wave~\cite{Loudon}:
%Type equation%
\begin{equation}\label{vectorpot}
 \boldsymbol{ \hat{A} } = \boldsymbol{ { A} }_{C}    + \boldsymbol{ \hat{A}}_{Q}.
\end{equation}
Here, $\boldsymbol{ \hat{A} }$ is the total vector potential. The quantities $\boldsymbol{ { A} }_{C}$ and $\boldsymbol{ \hat{A}}_{Q}$ refer to the classical vector potential and the quantized vector potential respectively. The quantized vector potential is given by~\cite{Loudon},
\begin{equation}\label{quantized vector potential}
    \boldsymbol{\hat{A}}_Q 
    = \sqrt{ \frac{2\pi}{  V  } } \sum_{\boldsymbol{k},\boldsymbol{\epsilon}}  \frac{1}{\sqrt{\omega_{k}}} \bigg[  \boldsymbol{\epsilon} e^{i\boldsymbol{k\cdot r}} \hat{a}_{\boldsymbol{k},\boldsymbol{\epsilon}} +  \boldsymbol{\epsilon}^* e^{-i\boldsymbol{k\cdot r} }
    { \hat{a}_{\boldsymbol{k},\boldsymbol{\epsilon}} ^\dagger  }\bigg]
\end{equation}
The symbols $\boldsymbol{\epsilon}$ and $\boldsymbol{k}$ refer to the unit polarization vector and wave vector of the photon respectively with $\boldsymbol{k} \cdot \boldsymbol{\epsilon} = 0$. Here, $\omega_{k}$ = \boldsymbol{$|k|$} c. The operators $\hat{a}_{\boldsymbol{k},\boldsymbol{\epsilon}}^\dagger$ and $\hat{a}_{\boldsymbol{k},\boldsymbol{\epsilon}}$ can create or annihilate a photon in mode $(\boldsymbol{k},\boldsymbol{\epsilon})$ respectively. The $V$ in the pre-factor refers to the volume of the region used to quantize the electromagnetic field modes. The quantity  $\boldsymbol{r}$ is the position vector and c is the speed of light in vacuum which is approximately 137.036 in a.u. It is to be noted that the final results are independent of the quantization volume $V$, because we consider the limit of an infinite volume.

The classical vector potential is modelled as a laser pulse with linear polarization. We choose the coordinate system such that the electric field only has a y-component and the X-ray pulse propagates in the x-direction. Our choice for this is given by the vector potential,
\begin{equation}\label{classicalvectorpotential}
    \boldsymbol{A}_C = \frac{E_C}{\omega_{in}} \cos\bigg[\omega_{in} (t - \frac{x}{c}) \bigg]  \exp\Bigg[\frac{(-2 \ln2 (t - \frac{x}{c})^2)}{t^2_{wid}} \Bigg]  \hat{y}    
\end{equation}
Here $E_C$, $\omega_{in}$, refers to the amplitude and the angular frequency of the incoming electric field respectively and 
$t_{wid}$ indicates the FWHM of the pulse intensity. It is to be noted that $\boldsymbol{A}_C$ is a function of $x$ and $t$ only. %A discussion on the values used for  $E_C$, $\omega_{in}$ and $t_{wid}$ is included in Sec. \ref{grid}.
%In our calculations, we use $E_C = 107.0$ a.u. and $\omega_{in} = 340$ a.u.

For the light-matter interaction, the Hamiltonian~\cite{Hamiltonianinteraction} is, 
\begin{equation}\label{full hamiltonian}
  \hat{H} =  \frac{(\boldsymbol{\hat{P}} + \boldsymbol{\hat{A}})^2}{2} +V(\boldsymbol{\hat{x} }) + \sum_{\boldsymbol{k},\boldsymbol{\epsilon}}\omega_{k}  \hat{a}_{\boldsymbol{k},\boldsymbol{\epsilon}}^{\dagger} \hat{a}_{\boldsymbol{k},\boldsymbol{\epsilon}}
\end{equation}  

Note that the exact form of the potential energy  $V(\boldsymbol{\hat{x} })$ is  discussed in Sec. \ref{application}. We use Eqs. (\ref{vectorpot}), (\ref{quantized vector potential}) and (\ref{full hamiltonian}) and separate out the terms with and without $\boldsymbol{\hat{A}}_Q  $. The terms with $\boldsymbol{\hat{A}}_Q $ are part of the perturbative correction. In this paper, we retain only the terms of first order in $\boldsymbol{\hat{A}}_Q $. One reason for this is that the higher order terms, give rise to two scattered photons and the Lamb shift, both of which are beyond the scope of this paper.
Thus, our unperturbed Hamiltonian is,
\begin{equation}
  \hat{H}^{(0)} =  \frac{(\hat{\boldsymbol{P}}+{\boldsymbol{A}}_C)^2}{2}  +  V(\boldsymbol{\hat{x} })  + \sum_{\boldsymbol{k},\boldsymbol{\epsilon}}\omega_{k}  \hat{a}_{\boldsymbol{k},\boldsymbol{\epsilon}}^{\dagger} \hat{a}_{\boldsymbol{k},\boldsymbol{\epsilon}}  %\frac{e^2 \boldsymbol{A}_C^2 }{2}  +  \boldsymbol{A}_C %\cdot\boldsymbol{\hat{P}}  
\end{equation}
\\
The perturbation term is,
\begin{equation}
 % \hat{H}^{ (1) } =   \boldsymbol{A}_C \cdot \boldsymbol{\hat{A}}_Q %+   \boldsymbol{\hat{A}}_Q \cdot \boldsymbol{\hat{P}}    
  \hat{H}^{ (1) } = ( \boldsymbol{\hat{P}}  + \boldsymbol{A}_C) \cdot \boldsymbol{\hat{A}}_Q
\end{equation}
The wave function is expanded in the Fock basis based on the number of scattered photons. Therefore our wave function ansatz is as follows,
\begin{equation}
    \ket{\psi_{total} } = \psi^{(0)} (\boldsymbol{r},t) \ket{0} + \sum_{\boldsymbol{k},\boldsymbol{\epsilon}} \psi_{\boldsymbol{k},\boldsymbol{\epsilon}} ^{(1)}(\boldsymbol{r},t) e^{-i \omega_{k} t} {\hat{a}_{\boldsymbol{k},\boldsymbol{\epsilon}} }^{\dagger}  \ket{0}
\end{equation}
where $\ket{0}$ refers to the vacuum state of the photon in Fock space. The ansatz is adequate because the first term describes an electron interacting with a classical EM field  without any scattered photons. The wave function of this electron is given by $\psi^{(0)} (\boldsymbol{r},t)$. The second term describes the presence of a scattered photon. The quantity $\psi ^{(1)}_{\boldsymbol{k},\boldsymbol{\epsilon}}(\boldsymbol{r},t)$ is
the probability amplitude at time t, for a photon to scatter into momentum $\boldsymbol{k}$ and polarization $\boldsymbol{\epsilon}$ and the electron to be found at position $\boldsymbol{r}$.

Given the Hamiltonian and the wave function ansatz, we proceed with the TDSE retaining only the terms up to first order in perturbation and separating out the equations based on the number of scattered photons. For no scattered photons,
\begin{equation} \label{eq_psi0}
   i \frac{\partial \psi ^ {(0)} }{\partial t} -  \hat{H}_{C}\psi ^{ (0) } = 0
\end{equation}
where,
\begin{equation}
    %\hat{H}_{C} \overset{\underset{\mathrm{def}}{}}{\cong}
            %\hat{H} ^ {(0)} -  \omega \hat{a}^{\dagger} \hat{a}
    \hat{H}_{C} =
            \hat{H} ^ {(0)} -  \sum_{\boldsymbol{k},\boldsymbol{\epsilon}}\omega_{k}  \hat{a}_{\boldsymbol{k},\boldsymbol{\epsilon}}^{\dagger} \hat{a}_{\boldsymbol{k},\boldsymbol{\epsilon}}    
\end{equation}
Note that $\hat{H}_{C}$ appears in Eq. (\ref{eq_psi0}) because $\psi^{(0)}$ is defined as the wave function of an electron interacting with a classical EM field.
%\begin{widetext} NOTE: split allows \\ but align does not. AVOID Eqnarray.
For 1 scattered photon we get,
\begin{equation} \label{eqn_psi1}
\begin{split} 
  i \frac{\partial \psi ^ {(1)}_{\boldsymbol{k},\boldsymbol{\epsilon}} }{\partial t} - \hat{H}_{C} \psi ^{ (1) }_{\boldsymbol{k},\boldsymbol{\epsilon}} =  &\sqrt{\frac{2\pi}{ V\omega_{k}} } e^{-i\boldsymbol{k\cdot r} }  e^{i\omega_{k} t }    \\    
  & \times \boldsymbol{\epsilon}^* \cdot (\boldsymbol{\hat{P}} + \boldsymbol{A}_C ) W(t) \psi ^{(0)}
\end{split}
\end{equation}
%\end{widetext}
where, 
\begin{equation}
    W(t) = e^{ - ( \frac{t}{ \tau}  ) ^ 8  } 
\end{equation}
The windowing function, W(t), adiabatically turns on the in-homogeneous term in Eq. (\ref{eqn_psi1}) only for the duration of the incident laser pulse, $t_{wid}$. This is done to find the ground state of the electron-photon coupled system. This also prevents the unphysical emission of photons, that would occur if the interaction between the electron and quantized photons was instantaneously turned on. Note that the function should be smooth to avoid encountering the Gibbs Phenomenon~\cite{Gibbs}. The choice of $\tau$ is determined by the duration of the pulse. The results of the calculation do not depend on $\tau$ as long as $\tau > 3.2~t_{wid}$ approximately. Another competing consideration is that, $\tau$ should be as small as possible to ensure that we only need to solve the TDSE for a short duration.  In our calculations we chose $\tau \sim 3.2 ~ t_{wid} $.  
%Here $t_{mid}$ is half the time-duration of the each computational %run and $tau = t_{mid} /1.57 $. 
The results do not depend on the specific choice of the windowing function as long as it is a smooth function which attains a value of 1, only during the duration of the incoming pulse.  

A modification of the procedure developed in this subsection is considered in Sec.\ref{eecorrelation} where the results of a two-electron calculation is discussed to probe electron-electron correlation effects in 2D. 
%In Sec.\ref{eecorrelation}, we modify the method developed in this section slightly to perform two-electron calculations and study electron-electron correlation effects in %2D.  
%\subsection{Modelling the vector Potential}
%[Insert Figure diffCross vs k for some thetas ]

%\begin{figure}
%\resizebox{80mm}{!}{\includegraphics{Test/130_2C_2D_final.eps}}
%\caption{\label{135-2C-2D}
%Scattering Probability as a function of wavenumber for NLC in 2D. %For a photon of incoming $\omega = 340$,the expected K$_{NLC}$ is %4.556 (a.u.). The peak is exactly at this value.
%}
%\end{figure}

\subsection{Relativistic Correction - ($\boldsymbol{P}+  \boldsymbol{A})^4$ terms} \label{Rel-correction} 
Here, we demonstrate how a relativistic correction may be implemented. We do this by considering the next higher order term in mechanical momentum and re-deriving the expressions in Eqs. (\ref{eq_psi0}) and (\ref{eqn_psi1}). A careful consideration of the non-commuting terms in ($\boldsymbol{\hat{P}}+  \boldsymbol{\hat{A}})^4$  is required to derive the new equations.
For no scattered photons, 
\begin{equation} \label{Rel_eq_psi0}
    i \frac{\partial \psi ^ {(0)} }{\partial t} - \hat{H}_{C}\psi ^{ (0) } =  - \frac{1}{8c^2}( \boldsymbol{\hat{P}} + \boldsymbol{A}_C  )^4  \psi ^{ (0) }
\end{equation}
%\begin{widetext}
For 1 scattered photon,
%\end{widetext}
\begin{equation} \label{Rel_eqn_psi1}
\begin{split}
   i \frac{\partial \psi ^ {(1)} _{\boldsymbol{k},\boldsymbol{\epsilon}} }{\partial t} - \hat{H}_{C} \psi ^{ (1) }_{\boldsymbol{k},\boldsymbol{\epsilon}} = &- \frac{1}{8c^2}( \boldsymbol{\hat{P}} + \boldsymbol{A}_C  )^4    \psi ^{ (1) }_{\boldsymbol{k},\boldsymbol{\epsilon}} \\
    &+  \sqrt{\frac{2\pi}{ V\omega_{k}  } }\boldsymbol{\epsilon}^* \cdot \bigg[  e^{-i\boldsymbol{k\cdot r} }e^{i\omega_{k} t }( \boldsymbol{\hat{P}} + \boldsymbol{A}_C  )\\ 
    & -\frac{1}{2c^2} e^{i\omega_{k} t } \hat{\boldsymbol{G}} \bigg]\psi ^{(0)}  
\end{split}
\end{equation}
where,
\begin{equation}
\begin{split}
   \hat{\boldsymbol{G}}  &=
    \big[e^{-i\boldsymbol{k\cdot r} } ( \boldsymbol{\hat{P}} + \boldsymbol{A}_C)^3 \big]  \\ 
    & + \big[ ( \boldsymbol{\hat{P}} + \boldsymbol{A}_C)^3 e^{-i\boldsymbol{k\cdot r} }\big]  \\
    & + \big[ ( \boldsymbol{\hat{P}} + \boldsymbol{A}_C  )^2 e^{-i\boldsymbol{k\cdot r} }( \boldsymbol{\hat{P}} + \boldsymbol{A}_C  ) \big]  \\
    & + \big[ ( \boldsymbol{\hat{P}} + \boldsymbol{A}_C  ) e^{-i\boldsymbol{k\cdot r} }( \boldsymbol{\hat{P}} + \boldsymbol{A}_C  )^2 \big]
\end{split}
\end{equation}

The above equations are a simple way in which relativistic corrections can be implemented. An alternative, more sophisticated approach would be to use the relativistic Schr\"odinger equation~\cite{Lindblom}. However, there is no need for such an approach given the results in Sec. \ref{freeelectron}. If the fields were a few orders of magnitude higher, there would be a need for a more sophisticated treatment of relativistic corrections~\cite{Forre}.

\subsection{Differential cross section}
The probability for a photon to scatter with momentum $\boldsymbol{k}$ and polarization $\boldsymbol{\epsilon}$ is
 \begin{equation} \label{scatteringprobability}
     P_ {\boldsymbol{k},\boldsymbol{\epsilon}} = \int_v {\psi_ {\boldsymbol{k},\boldsymbol{\epsilon}}^{(1)}}^* \psi_ {\boldsymbol{k},\boldsymbol{\epsilon}}^{(1)} d^{n}r
 \end{equation}
Here $d^{n}r$ refers to the volume element in $n$ dimensions.

The method described in Sec.~\ref{derivingnon-homogenoeous} automatically leads to a spread in the scattered photon momentum because the incoming field is not strictly monochromatic but rather a pulse. The amount of the spread in scattered photon momentum is determined by the width of the chosen laser pulse. Since there is a momentum spread in the scattered X-ray, the differential cross section for a given scattering angle is a summation over all possible magnitudes of scattered photon momentum.
The total 1-photon cross section in 3D is given by,

 \begin{equation}
     \sigma^{(1)} = \sum_{\boldsymbol{k},\boldsymbol{\epsilon} } \frac{P_{\boldsymbol{k},\boldsymbol{\epsilon}} }{\text{(number\ of\ photons/area)}} 
 \end{equation}
 
where \cite{Loudon} , 
    \begin{equation}
         \sum_{\boldsymbol{k} }  \longrightarrow \frac{V}{(2\pi)^3} \int d^3k
    \end{equation}
    
and, 
\begin{equation}
    \frac{\text{number\ of\ photons}}{\text{area}} = \frac{\int I dt }{\omega_{in}}
\end{equation}
Here $I$ refers to the intensity of incoming field. It is to be noted that the incoming pulse is assumed to be quasi-monochromatic.
This leads to the definition of differential cross section:
    \begin{equation}
        \dv{\sigma}{\Omega}^{(1)} =
        \frac{V \omega_{in} }{(2\pi)^3}\frac{\int \sum\limits_{ \boldsymbol{\epsilon} } P_ {\boldsymbol{k},\boldsymbol{\epsilon}} k^2dk }{\int I dt } 
    \end{equation}
     Here $V$ is the quantization volume. There exists a factor of 1/$V$ in $P_{\boldsymbol{k},\boldsymbol{\epsilon}}$ which cancels out the $V$ in the numerator. Note that $\omega_{in}$ refers to the angular frequency of the incoming electric field [Eq. (\ref{classicalvectorpotential})].  
     
The 2-photon cross section in 3D has been defined in multiple ways~\cite{2photoncrosssection1986,faisalmultiphoton}. Here we define it so that the SI units would be $m^2/(W/m^2)$.
     \begin{equation}
         \sigma^{(2)} = \omega_{in}   \frac{\sum\limits_{\boldsymbol{k},\boldsymbol{\epsilon} }P_ {\boldsymbol{k},\boldsymbol{\epsilon}} }{\int I^2 dt } 
     \end{equation}
Therefore the differential cross section would be, 
     \begin{equation}
         \dv{\sigma}{\Omega}^{(2)} = \frac{V  \omega_{in} }{(2\pi)^3}\frac{\int \sum\limits_{ \boldsymbol{\epsilon} } P_ {\boldsymbol{k},\boldsymbol{\epsilon}} k^2dk }{\int I^2 dt }
     \end{equation}
In both the 1-photon and 2-photon differential cross-sections, we calculate these integrals with respect to $k^2dk$ by doing a Gaussian fit for the plots of $P_{\boldsymbol{k},\boldsymbol{\epsilon}}$ vs k and then performing an integral of the Gaussian function. The differential cross sections obtained from the 3D calculations do not have any adjustable parameters.    

The exact factors to obtain the differential cross length from the scattering probability in 2D are not well defined. Therefore, we obtain this factor by scaling our differential cross sections to get an overall fit with the analytical free electron results~\cite{QED2plus1,KB}.

\begin{figure}
\resizebox{80mm}{!}{\includegraphics{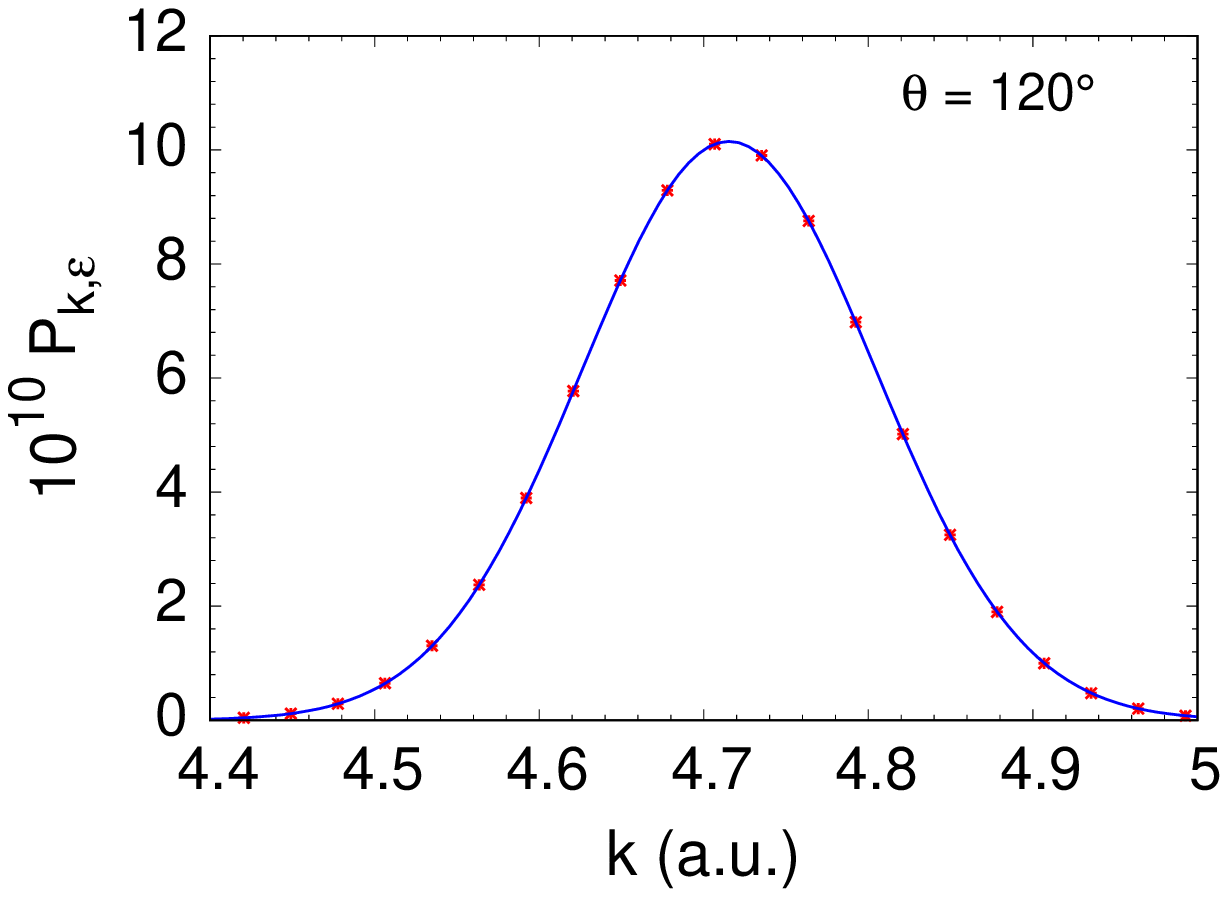}}
\resizebox{80mm}{!}{\includegraphics{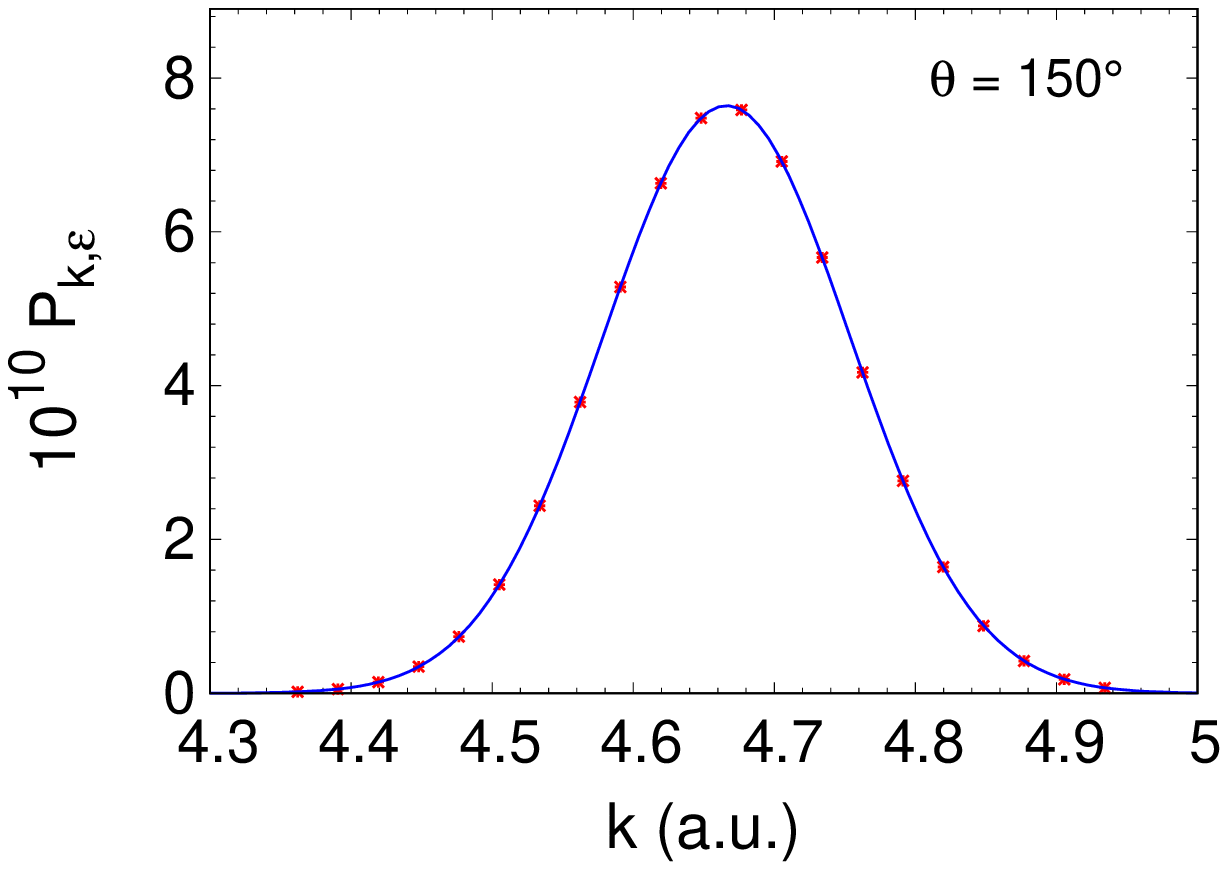}}
\caption{\label{Gaussianfits2D}
Scattering probability $P_{\boldsymbol{k},\boldsymbol{\epsilon}}$ as a function of scattered photon momentum $k$ for non-linear Compton calculations in 2D for an angle of 120 and 150 degrees respectively for a free-electron. The red points indicate the results of the numerical calculation and the blue line indicates a Gaussian fit. The calculations were performed over an equal number of $k$ values on either side of the theoretical value. Note that the peaks are at the expected non-linear Compton momentum [Eq. (\ref{NLC_freq})]. This calculations was done with $E_C$ = 107 a.u. , $\omega_{in}$ = 340 a.u. and $t_{wid}$ = 0.125 a.u.
}
\end{figure}
\subsection{ Solving TDSE } \label{solvingTDSE}
%for $ \psi^{0,1}(\boldsymbol{r},t)$
We solve the TDSE using a Cartesian co-ordinate system with the wave function represented on a grid of points. The values of the grid parameters are specified in Sec. \ref{grid}. For the kinetic energy operator in the Hamiltonian, we use a three-point central difference formula. The TDSE for $\psi ^{(0)}(\boldsymbol{r},t)$ [Eq. (\ref{eq_psi0})] is solved using the leap-frog method~\cite{numericalrecipes}. We choose the leap-frog method for two reasons: first, it preserves unitarity and second, it leads to converged results which is discussed in detail in Sec. \ref{grid}. The leap-frog method involves computing the wave function which is two time-steps ahead of the current wave function, using the wave function at the current time-step and the wave function at the intermediate time. The second order Runge-Kutta method is used to obtain the value of the wave function at the first time-step which is required for the leap-frog approach. At every time instance, we simultaneously solve for $\psi ^{(1)}_{\boldsymbol{k,\epsilon}}(\boldsymbol{r},t)$ for a range of scattered photon momenta centered around the Compton momentum or the Brown and Kibble prediction [Eq. (\ref{NLC_freq})] for the linear Compton or non-linear Compton respectively.  The plot of $P_{\boldsymbol{k},\boldsymbol{\epsilon}}$ [Eq. (\ref{scatteringprobability})] as a function of scattered photon momentum $k$ is a Gaussian curve (see Fig. \ref{Gaussianfits2D}) to a good approximation. 

In all our calculations unless otherwise stated, we use electric field $E_C$ = 107 a.u. and angular frequency $\omega_{in} = 340$  a.u. for the incident laser pulse. In SI units, these values correspond to an electric field of $ \sim 5$ x $10^{13}$ V/m  and an intensity of $\sim 3$ x $10^{24}$ W/m\textsuperscript{2}. The chosen angular frequency corresponds to an incoming photon energy of about 9.25 keV. These values belong to the range used in the experiment by Fuchs et al. \cite{Fuchs}.
\subsection{Grid and other numerical parameters} \label{grid}
In our calculations, convergence is measured in two ways, by calculating the area under $P_{\boldsymbol{k},\boldsymbol{\epsilon}}$ vs k plots and by calculating the change in the peak position of the scattering probability. For all the calculations except Sec. \ref{eecorrelation},  the change in this area with respect to change in grid-spacing or grid-size were under 2\%. The change in the peak position of scattering probability with respect to change in grid-spacing or grid-size was under 0.5\% . 

For the 2D free-electron calculations, a grid size of 400 X 400 with a grid-spacing of 0.1 a.u. in both x,y directions resulted in converged results. For the 3D calculations with Z=1, Z=2 a grid range of 400 x 400 x 400 with a grid spacing of 0.1 a.u. resulted in converged results. For Z=4 a grid range of 229 X 229 X 229  with a grid-spacing of 0.07 units resulted in converged results.

 The primary source of error in scattered photon momentum arises from the kinetic energy operator. The leading order error term is proportional to the square of the grid-spacing. For the case of non-linear Compton scattering from a bound electron at an angle of 130 degrees and for a grid-spacing of 0.07 a.u., the error is of the size of about 3\% of the non-linear Compton shift. This error is much smaller than the size of the anomalous shift observed in Ref. ~\cite{Fuchs} which is about 100\% of the Compton shift. In Sec. \ref{boundelectroncase}, we take our estimate for scattered photon momentum, k below this 3\% error by using Richardson's extrapolation to eliminate the leading order error term.  

 For $t_{wid}$, we use a range of 0.1 - 1 a.u. which corresponds to a pulse of duration $\sim 10^{-18}$s. The use of such a short pulse is justified because the results for differential cross section are found to be independent of the choice of $t_{wid}$. A small change in peak scattered momentum is observed for different pulse widths. The magnitude of this change is less than about 1\% of the momentum shift observed by Fuchs et al.~\cite{Fuchs}. Also, for the chosen range of $t_{wid}$, there is no reflection of the wave function from the walls, as the distance travelled by the wave packet of the electron is much smaller than the size of the grid.

\section{Application}\label{application}
\subsection{Free-electron Case}\label{freeelectron}
\begin{figure}
\resizebox{80mm}{!}{\includegraphics{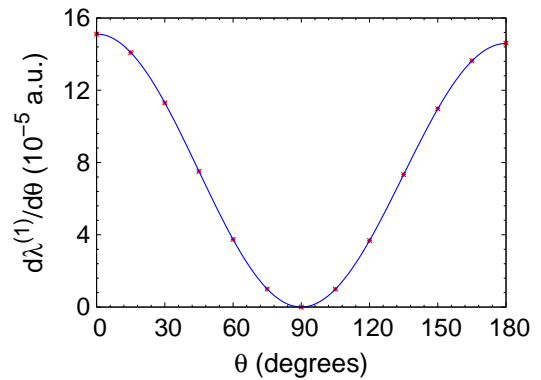}}
\caption{\label{QED2plus1}
Comparison of differential cross length as a function of angle subtended by the detector with the analogue of Klein-Nishina formula for 2D~\cite{QED2plus1} for linear Compton scattering. The red points are the results of the numerical calculation and the blue line represents the results from the analytical expression~\cite{QED2plus1}. The results of the numerical calculations in 2D were scaled by a single factor. This factor was chosen such that, overall, the numerical results fit well with the analytical results. The above calculations were done with the same parameters as Fig. \ref{Gaussianfits2D}.
}
\end{figure}

\begin{figure}
\resizebox{80mm}{!}{\includegraphics{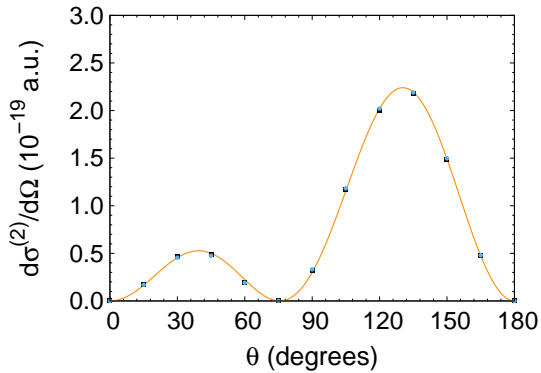}}
\caption{\label{KB_fit}
Comparison of differential cross section(length) as a function of scattering angle in 2D with the results of Brown and Kibble for non-linear Compton scattering. The blue points indicate the non-relativistic results obtained using Eqs. (\ref{eq_psi0}) and (\ref{eqn_psi1}). The black squares were obtained using the approach from Sec. \ref{Rel-correction}. The orange line indicates the result from Brown and Kibble. The results of the numerical calculations in 2D were scaled by a single factor. This factor was chosen such that overall the numerical results fit well with the analytical results. The above calculations were done with the same parameters as Fig. \ref{Gaussianfits2D}.
}
\end{figure}

We apply the method developed in Sec. \ref{Methods}, to a free electron interacting with a laser pulse in 2D and compare the results of our calculation with the equivalent of the Klein-Nishina formula in 2D~\cite{QED2plus1}. Note that the Klein-Nishina formula and its analogue in 2D are derived for monochromatic radiation. Since we employ a pulse, we evaluate the integral $D = \int\sum\limits_{ \boldsymbol{\epsilon} } P_ {\boldsymbol{k},\boldsymbol{\epsilon}}kdk$ to find a quantity proportional to the differential cross length for a given intensity and incoming frequency. 
We compute this quantity $D$ for different angles and compare this with the differential cross length from the QED-2+1 scheme~\cite{QED2plus1} and the differential cross section from Brown and Kibble~\cite{KB}. From this point in our discussions, we will refer to $D$ as differential cross length for convenience keeping in mind that the calculation has been scaled to match the analytical result. 

We plot the differential cross length we obtained for linear Compton as a function of angle and the results from QED-2+1~\cite{QED2plus1} in Fig. \ref{QED2plus1}. Upon comparison, we find that our calculated differential cross length agrees well with the free-electron analytical results.

Next, we compare the calculated 2D differential cross length (see Fig. \ref{KB_fit}) for non-linear Compton scattering with the analytical expression from Brown and Kibble~\cite{KB}. We also evaluate the differential cross lengths using the relativistic corrections developed in Sec. \ref{Rel-correction} for comparison. The procedure for scaling the differential cross length used previously, is employed here as well. According to Brown and Kibble~\cite{KB}, for non-linear Compton scattering, the frequency of the scattered photon using a non-relativistic approximation is given by, 
\begin{equation} \label{NLC_freq}
 {\omega} = \frac{n\omega_{in} }{1 + n\alpha^2 \omega_{in} (1 - cos \theta) }
\end{equation} 
Here, $n$ determines the order of the process, for example $n = 1$ for Compton scattering. The discussions in this paper are restricted to processes where $n \leq$ 2. The symbols $\omega$,  $\omega_{in}$ refers to the angular frequency of the scattered photon and incoming photon respectively and $\alpha$ is the fine-structure constant. 

It is important to note that the expression for differential cross section by Brown and Kibble was derived in 3D, but our calculations are for differential cross length in 2D. Upon comparison, we find that that our results are in good agreement with the Brown and Kibble results up to a constant factor. There is also no significant change (see Fig. \ref{KB_fit}) in the agreement with Brown and Kibble's result because of the relativistic correction discussed in Sec. ~\ref{Rel-correction}. Brown and Kibble had arrived at their results by solving the Dirac equation but our agreement with their results justifies the approximation with the TDSE.

It was found that the scattering probability for non-linear Compton exhibits a second-order dependence on the intensity of incoming EM field as expected and the scattering probability for Compton scattering exhibits a first-order dependence on the intensity of the incoming EM field. This behaviour was observed over at least 3 orders of magnitude (up to 1000 a.u.) in the electric field.

\subsection{Bound electron case} \label{boundelectroncase}

\begin{figure}
\resizebox{80mm}{!}{\includegraphics{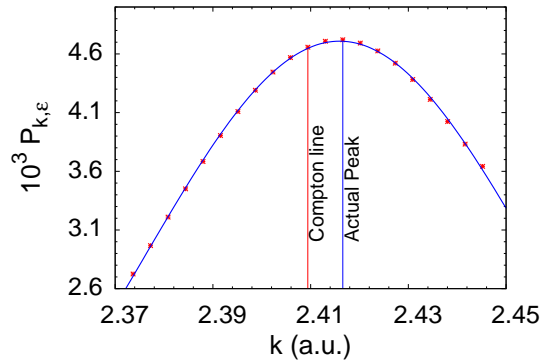}}
\caption{\label{highres1C}
The above plot was computed by solving the problem in 2D for $Z=4$, $a=0.1$ a.u., with a binding energy(BE) of 5.9593 a.u. at an angle of 130 degrees and $t_{wid} = 1$. It reveals the Compton defect in linear Compton scattering. The red vertical line and the blue vertical line indicates the expected peak (non-relativistic) and the actual peak respectively in the scattered photon momentum $k$. The red points indicate the results of the numerical calculation and the blue curve indicates a Gaussian fit.
}
\end{figure}
\begin{figure}
\resizebox{80mm}{!}{\includegraphics{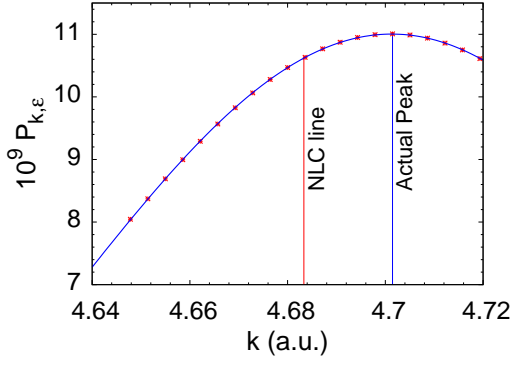}}
\caption{\label{highres2C}
The above plot was computed by solving the problem in 2D at an angle of 130 degrees for $Z=4$, $a=0.1$ a.u. leading to a binding energy of 5.9593 a.u. The red points indicate the results of the numerical calculation and the blue curve indicates a Gaussian fit. It reveals an analogue of the Compton defect in non-linear Compton scattering. The red vertical line and the blue vertical line indicates the expected peak (non-relativistic) and the actual peak respectively in the scattered photon momentum $k$. Here, $t_{wid} = 1$ .  
}
\end{figure}
Here we consider the case of bound electrons because of its relevance to Fuchs et al.~\cite{Fuchs}. Unlike the calculations for a free electron, here we adopt a 3D approach for the most part. It is to be noted that a 3D calculation can be done with relative ease for the case of a bound electron, as the grid needed for convergent solutions is smaller. Hence, it involves less memory and time computationally when compared to the case of a free electron. 

While the method developed in Sec.~\ref{Methods} allows for flexibility with respect to the choice of potential, to keep things simple a softcore Coulombic potential of the  following form is chosen: 
\begin{equation} \label{atomicpotential}
    V(\boldsymbol{r}) = \frac{- Z }{\sqrt{x^2 + y^2 + z^2 +a} }
\end{equation}
Here, $Z$ is equal to the effective nuclear charge seen by the electron in atomic units. By varying this, we can model the scattering from bound electrons of different binding energy(BE). The parameter $a$ is included to avoid the  singularity~\cite{Singularity2,Singularity3,Singularity4,singularity}  at the origin. While it is preferable to minimize the value of this parameter, there are constraints that arise from the grid-spacing. 

\begin{figure}
\resizebox{80mm}{!}{\includegraphics{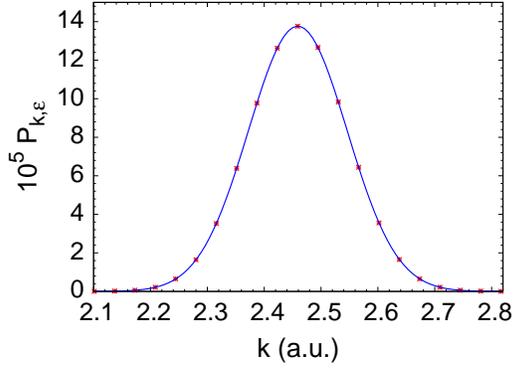}}
\caption{\label{Gaussianfit3Db1}
Scattering profile for Compton scattering for a bound electron in 3D at an angle of 60 degrees with $t_{wid} = 0.1$. The bound state of the electron is characterised by parameters $Z=4$, $a=0.1$ a.u. leading to a BE of 3.9496 a.u.  The red points indicate the results of the numerical calculation and the blue line indicates a Gaussian fit. 
}
\end{figure}
\begin{figure}
\resizebox{80mm}{!}{\includegraphics{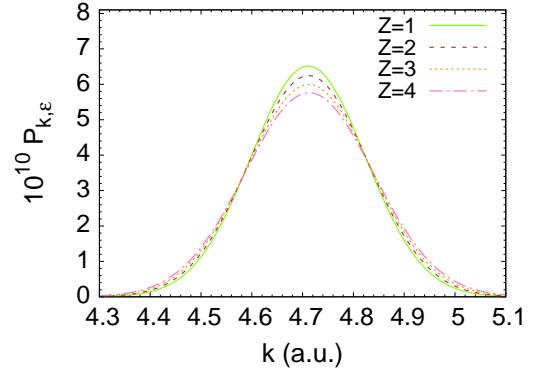} }
\caption{\label{Gaussianfit3Db2}
Scattering profile for non-linear Compton scattering for a bound electron in 3D at an angle of 120 degrees with $t_{wid} = 0.1$. The figure contains the Gaussian fits from bound states characterised by Z=1, 2, 3, 4 with binding energies (a.u.) 0.4037, 1.322, 2.5345, 3.9449 respectively. Here $a=0.1$, $E_C$ = 107 a.u., $\omega_{in}$ = 340 a.u. In the experiment by Ref.~\cite{Fuchs}, the peak was observed at a $k$ value of  $\sim$ 4.5 a.u. , but from the calculations, the bound nature of electron doesn't appear to have altered the peak scattered momentum from the free electron value.
}
\end{figure}

With this potential, we proceed as per Sec. \ref{Methods} and obtain the scattering probability,  $P_{\boldsymbol{k},\boldsymbol{\epsilon}}$ (see  \crefrange{highres1C}{Gaussianfit3Db2}).
\begin{figure}
\resizebox{80mm}{!}{\includegraphics{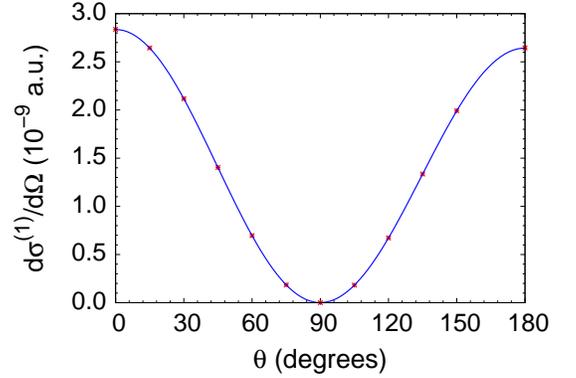}}
\caption{\label{KNbound}
Comparison of the differential cross section for Compton scattering from a bound electron as a function of scattering angle with the results of Klein-Nishina formula. The red points indicate the results of the numerical calculation and the blue line indicates the results of Klein-Nishina formula. The above calculations were done with the same parameters as in Fig. \ref{Gaussianfit3Db1}. All numerical calculations in 3D were done with no adjustable parameters. 
}
\end{figure}
From the scattering probability calculations for non-linear Compton scattering for different bound state parameters, two things  should be noted. First, there \textit{is} a momentum shift, albeit an insignificant one when compared to the shift measured by Fuchs et al~\cite{Fuchs}. Second, Ref.~\cite{Fuchs} measured a redshift while the simulations show a blue-shift.
While the additional shift in Compton wavelength has been well documented and studied~\cite{bergestorm,Bloch}, interestingly we find that a similar shift occurs in non-linear Compton as well. %This calculated shift in scattered momentum for non-linear Compton scattering is of the order  $\sim 10^{-3}$ atomic units. The shift in scattered photon momentum observed by Fuchs et al.~\cite{Fuchs} is of the order of 0.1 atomic units in the opposite direction.

\begin{figure}
\resizebox{80mm}{!}{\includegraphics{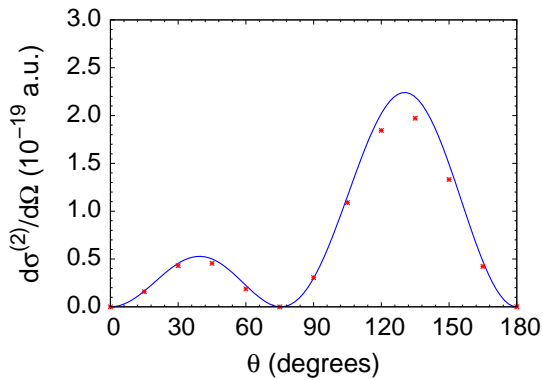} }
\caption{\label{KBfit3D}
Comparison of the differential cross section as a function of scattering angle for non-linear Compton scattering from bound electrons with the Brown and Kibble's free-electron result. The bound electron is characterised by parameters $Z=4$ and $a=0.1$ a.u. with a BE of 3.9496. The red points are a result of the numerical calculations in 3D while the blue line indicates the results of Brown and Kibble. The above calculations were done with the same parameters as Fig. \ref{Gaussianfit3Db2}. All numerical calculations in 3D were done with no adjustable parameters.\\
}
\end{figure}

We calculate the differential cross section for Compton and non-linear Compton as a function of angle for a bound electron. When we compare the calculated linear Compton differential cross section with the Klein-Nishina formula~\cite{KN}, we find excellent agreement (see Fig. \ref{KNbound}) despite it being a bound electron. Upon comparing the non-linear Compton differential cross section with Brown and Kibble's result~\cite{KB}, we find a general agreement (see Fig. \ref{KBfit3D}). However, the calculated differential cross section for angles between 120 degrees and 150 degrees exhibit about 10 percent discrepancy. A part of this discrepancy arises from the fact that the Brown and Kibble formula used was non-relativistic and therefore is missing factors of ($\frac{\omega_{k}}{\omega_{in}}$). The differential cross section for Compton scattering from Kibble and Brown is also missing these factors which were included in the Klein-Nishina formula. Therefore, it is difficult to determine the amount of error that originates from the numerical calculation and the amount that originates from the electrons being bound. These calculations were done for a range of values for $Z$ and $a$ and the results were found to be approximately the same. 

It is to be noted that the results of Krebs et al. \cite{Santra} are in terms of double differential cross section and their double differential cross section has an extra frequency factor. Upon finding the area under their curve for double differential cross section by approximating it as a Gaussian and after accounting for differences in frequency,  we find that our results are of the same order as theirs and agree to within a factor of $\sim$2. This comparison is approximate because the estimate for the area is crude due to the limited number of data points in the results of Ref.~\cite{Santra}. 

For all the calculations, the polarization of the scattered photon was in the same plane as that of the plane of polarization of the incoming photons. When the polarization of the scattered photon was chosen to be perpendicular to the plane of polarization of the incoming photons, the scattering probabilities were found to be more than 6 orders of magnitude smaller, for the case of non-linear Compton scattering. 

For calculating the additional shifts(defect) in $k$, we first numerically calculate the average $k$ instead of obtaining the peak momentum from the Gaussian fit. When the polarization of scattered photon is in the plane of polarization of the incoming photons, 
\begin{equation}
    k_{avg} =  \frac{\sum\limits_k k P_k} {\sum\limits_k P_k}
\end{equation}
Here $k_{avg}$ is the estimate for the scattered photon momentum that we use to calculate the defect, with respect to the theoretical non-relativistic free electron prediction for both linear and non-linear Compton scattering. Because the X-rays in the calculation have a Gaussian time dependence, the final momentum distribution is the convolution of the infinite resolution distribution with a Gaussian. The average of the final k is unchanged by the convolution because the Gaussian is a symmetric function while the peak value does slightly shift with the $t_{wid}$. In the calculations here, the scattering probability falls off slower than a Gaussian distribution for $k$ values far from that for free electron linear and non-linear Compton scattering which leads to small shifts. The underlying cause for this lies in the nature of the Compton profile of the bound electron. Following this, Richardson's extrapolation method~\cite{numericalrecipes} is used to obtain an estimate for the defect in scattered photon momentum after accounting for the numerical error from the grid-spacing to the leading order. For the cases of $Z$ = 1, 2, 3 and 4 with $a$ = 0.1, the defects were found to be of the size of $\sim 10^{-3}$ a.u. in $k$ which corresponds to an energy of about a few eV. It was found that the size of the defect increases with the binding energy of the electron. The defect was also found to be independent of the incident field over the range 1-110 atomic units of electric field amplitude.  

Let $k_{final}$ and $k_{initial}$ be the peak scattered momentum of the outgoing Xray photon and the peak momentum of the incoming Xray photons respectively. For the case of non-linear Compton scattering from a free electron at an angle of 120 degrees,  $k_{final} - k_{initial}$ $\sim$ - 0.25 a.u. From the experiment~\cite{Fuchs}, $k_{final} - k_{initial}$ $\sim$ - 0.5 a.u. From our bound electron calculations, we find that $k_{final} - k_{initial}$ $\sim$ - 0.25 a.u. but there is a small blueshift correction to this which is of the size $\sim$ + $10^{-3}$ a.u.  From these results, it is evident that the bound nature of the electron cannot explain the anomalous shift observed in Ref.~\cite{Fuchs}.

\subsection{Electron-electron correlation effects} \label{eecorrelation}
We examine if electron-electron interaction effects could contribute to the redshift in the non-linear Compton scattering. This can be done by a simple extension of the procedure developed in Sec. \ref{Methods}. The Hamiltonian is modified to include the mechanical momentum from each electron and an interaction potential is introduced. 
\begin{figure}
\resizebox{80mm}{!}{\includegraphics{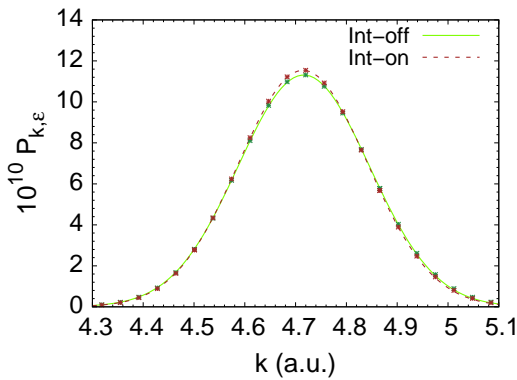} }
\caption{\label{Gaussianfit_ee_2C}
 Scattering Profile for non-linear Compton scattering for an angle of 130 degrees. The curves represent Gaussian fits while the points are the result of the numerical calculation. The dark green points and the green curve represents the case with the electron-electron interaction turned off and the brown points and the brown dotted line indicates the case with the interaction turned on. Here $Z$ = 4, $a$ = 0.1, $E_C$ = 107 a.u., $\omega_{in}$ = 340 a.u., $t_{wid}$ = 0.1 a.u. 
}
\end{figure}

The modified Hamiltonian is given by,
\begin{equation}\label{ee hamiltonian}
\begin{split}
  \hat{H} =  & \frac{(\boldsymbol{\hat{P}}_1 + \boldsymbol{\hat{A}}(\boldsymbol{r}_1))^2}{2}+ \frac{(\boldsymbol{\hat{P}}_2 + \boldsymbol{\hat{A}}(\boldsymbol{r}_2))^2}{2} + V(\boldsymbol{r}_1) +  V(\boldsymbol{r}_2) \\
  &+ \frac{1}{\sqrt{(x_1 - x_2)^2 + (y_1 - y_2)^2  + a}}  \\
  &+ \sum_{\boldsymbol{k},\boldsymbol{\epsilon}}\omega_{k}  \hat{a}_{\boldsymbol{k},\boldsymbol{\epsilon}}^{\dagger} \hat{a}_{\boldsymbol{k},\boldsymbol{\epsilon}}
  \end{split}
\end{equation}
Here $\boldsymbol{r}_1$ and $\boldsymbol{r}_2$ refer to the position vectors of the electrons and $V(\boldsymbol{r}_1)$ and $V(\boldsymbol{r}_2)$ is the 2D equivalent of the expression in Eq. (\ref{atomicpotential}). The wave function ansatz remains the same except that the quantities $\psi^{(0)}$ and  $\psi ^{(1)}_{\boldsymbol{k},\boldsymbol{\epsilon}}$  are now functions of both $\boldsymbol{r}_1$ and $\boldsymbol{r}_2$ along with time t.

With this approach, the calculations have to be restricted to 2D because of the time and space required to handle the problem computationally. Restricting the calculation to 2D is reasonable given that there was not any significant difference in the 2D and 3D results from Sec. \ref{freeelectron} and Sec. \ref{boundelectroncase} respectively. 

The same numerical procedure discussed in Sec. \ref{solvingTDSE} is used to obtain the scattering probability $P_{\boldsymbol{k},\boldsymbol{\epsilon}}$ as a function of scattered photon momentum $k$. A comparison of the calculation with and without the electron-electron interaction does not indicate any significant change (Fig. \ref{Gaussianfit_ee_2C}). 

This calculation is performed with a grid-spacing of 0.14 a.u. and therefore it is not converged to the same extent as the previous calculations. In single bound electron calculations in 2D and 3D, there is no substantial change in the nature of our results as the grid-spacing is decreased from 0.2 to 0.07 atomic units. We extrapolate from this trend and argue that the electron-electron correlation effects are unlikely to be cause of the redshift observed in the experiment by Fuchs et al.~\cite{Fuchs}. 

\subsection{Semi-Compton process}\label{semicompton}
We consider a process where a bound electron absorbs an incoming photon and the now-ionised electron scatters another incoming photon inelastically to give rise to a photon of frequency $\sim$ $2\omega_{in}$. The electron ends up being re-captured by the atom during the process. This process should manifest itself in the calculations if the grid-spacing was decreased enough to access the  energy range in the continuum of the ionised electron. When the bound electron absorbs a photon, it gains a momentum of $\sim$ 26 a.u.  This would not be represented in a grid with a spacing of 0.1 a.u., hence we consider a grid-spacing of 0.02 a.u. 

We resort to a 2D calculation to probe such a fine grid. The calculations do not reveal any significant difference in the scattering profile. We also consider the effect of binding energy on this scattering profile by decreasing the parameter $a$ in the potential. We do not find any significant effect beyond the Compton defect discussed in Sec. \ref{boundelectroncase} which is atleast 2 orders of magnitude smaller than the shift observed by Fuchs et al. \cite{Fuchs}.
\section{Conclusion and Summary}
We described a method to numerically calculate the linear and non-linear Compton effect for free or bound electrons. The results from the calculation can be used to determine whether the bound nature of the electrons caused the anomalous frequency shift observed in the experiment by Fuchs et al.~\cite{Fuchs}. To justify the approximations we compared our free-electron results with the analytical expressions available for differential cross sections of Compton~\cite{QED2plus1,KN} and non-linear Compton scattering~\cite{KB}. We found excellent agreement in those cases.

We employed a Coulombic interaction potential to model bound electrons and obtained their differential cross sections for Compton and non-linear Compton scattering. Despite the electrons being bound, the calculations for the differential cross section agreed with the Brown and Kibble results. The calculations did not exhibit a redshift in the wavelength of the scattered photon, in disagreement with the experiment~\cite{Fuchs} but in agreement with the calculations of Krebs et al.~\cite{Santra}. For bound electrons, we also found the small expected blue shift in the case of Compton scattering and interestingly a blue shift in the case of non-linear Compton scattering as well. Our calculations support the conclusion in Ref.\cite{Santra} that it is not the bound character of the electron that is causing the anomalous frequency shift seen in Fuchs et al.~\cite{Fuchs}.

The role of electron-electron correlation effects on the redshift was explored by doing a two-electron calculation in 2D. The results of the calculation did not indicate the presence of the redshift in Ref.~\cite{Fuchs}.  Following this, we considered the case of a semi-Compton process where linear Compton-scattering occurs off of an ionised electron with the electron getting re-captured. This could give rise to a photon of frequency of $\sim$ $2\omega_{in}$. A calculation accounting for this process, did not exhibit a redshift similar to the one observed in the experiment by Fuchs et al. \cite{Fuchs}. No calculations have yet been able to reproduce the shift observed in Ref.~\cite{Fuchs}.
\section{Acknowledgements}
This material is based upon work supported by the US Department of Energy, Office of Science, Basic Energy Sciences, under Award No. DE-SC0012193. We thank the Science IT at Department of Physics, Purdue University for their assistance. We thank D.A. Reis and P.H. Bucksbaum for our discussions about their experiment. We are also grateful to D. Krebs, D.A. Reis and R. Santra for providing a pre-print of their work. A.V. thanks X. Wang and T. Seberson for discussions on computational issues.

% \begin{figure}
% \end{figure}

% Surround figure environment with turnpage environment for landscape
% figure
% \begin{figure}
% \end{figure}

% tables should appear as floats within the text
% The ruledtabular enviroment adds doubled rules to table and sets a
% reasonable default table settings.
% Use the table* environment to get a full-width table in two-column
% Add \usepackage{longtable} and the longtable (or longtable*}
% environment for nicely formatted long tables. Or use the the [H]
% placement option to break a long table (with less control than 
% \begin{table}%[H] add [H] placement to break table across pages
% \begin{ruledtabular}
% Lines of table here ending with \\
% \end{tabular}
% \end{ruledtabular}
% \end{table}

% Surround table environment with turnpage environment for landscape
% table
% \begin{table}
% \begin{ruledtabular}
% \end{tabular}
% \end{ruledtabular}
% \end{table}

\appendix*

%\section{}

%\begin{acknowledgments}
%\end{acknowledgments}

%\section{References}
\bibliography{References.bib}
%Blablabla said Nobody ~\cite{Nobody06}.
%1) Fuchs et.al
%2) Kibble and Brown 
%3) Santra et. al
%4) QED 2+1 
%5) 

\end{document}